\documentclass[12pt]{article}
\usepackage{latexsym}
\let\b=\beta\let\d=\delta

\let\r=\rho

\def\ket#1{|#1\rangle}

\def\IR{\relax{\rm I\kern-.18em R}}

\newcommand{\SL}[0]{{\rm SL}(2,\IR)}

\newcommand{\be}{\begin{equation}}
\newcommand{\ee}{\end{equation}}
\newcommand{\beq}{\begin{equation}}
\newcommand{\eeq}{\end{equation}}
\newcommand{\bea}{\begin{eqnarray}}
\newcommand{\eea}{\end{eqnarray}}

\newcommand{\del}{\partial}

\newcommand{\nbox}{{\,\lower0.9pt\vbox{\hrule \hbox{\vrule height 0.2 cm \hskip 0.2 cm \vrule height 0.2 cm}\hrule}\,}}

\begin{document}

\rightline{hep-th/0108001}
\rightline{RUNHETC-2001-21}
\vskip 1cm
\centerline{\Large \bf Boundary States for D-branes in $AdS_3$}
\vskip 1cm

\renewcommand{\thefootnote}{\fnsymbol{footnote}}
\centerline{{\bf \large Arvind
Rajaraman\footnote{arvindra@muon.rutgers.edu}
and Moshe Rozali\footnote{rozali@muon.rutgers.edu}}}
\vskip .5cm
\centerline{\it Serin Laboratory, Rutgers University}
\centerline{\it Piscataway, NJ 08854, USA}
\vskip .5cm

\setcounter{footnote}{0}
\renewcommand{\thefootnote}{\arabic{footnote}}

\begin{abstract}
We construct boundary states representing D-strings in $AdS_3$.
These wrap twisted conjugacy classes of $\SL$, and
the boundary states are therefore based on continuous
representations only.
We check Cardy's condition and find a consistent open string spectrum.
The open string spectrum on all the D-branes is the same.
\end{abstract}

\section{Introduction}
Perturbative string theory on $AdS_3$, which is isomorphic
to the $\SL$ group manifold, is considerably more
complicated than string theory on compact group manifolds.
(For previous work on $\SL$, see \cite{
Petropoulos:1999me,Petropoulos:1999nc,
noghost,Giveon:1998ns,Gawedzki:1991yu,
deBoer:1998pp,
Teschner:2000ug,
Kato:2000tb}.)
Some of the subtleties of the closed string spectrum 
were worked out in \cite{mo}, where a proposal for 
the closed string spectrum was proposed, and checked
by an explicit computation of the partition function.

In this paper, we perform an analogous analysis
for open strings in $AdS_3$ i.e. strings 
ending on D-branes in  $AdS_3$. D-branes were discussed from
a semiclassical point of view in \cite{Stanciu:1999nx,
Figueroa-O'Farrill:2000ei,bachas}. We will
present here a conformal field theory description of
these branes as boundary states.

The problem we have to face in constructing
boundary states is that their interactions
are divergent. This is because unitary representations
of $\SL$ are all infinite dimensional. 
Characters of these representations tend to be ill defined.

On the other hand, the simple reason why D-brane interactions
are ill-defined is because all the branes
we consider have an infinite volume. We show that
in an appropriate system of calculation, all divergences
can indeed be understood as volume divergences, and as such
can be easily regularized. 

Once the overlap of these boundary states is regularized,
we can then check Cardy's condition. In particular,
we can compare the open string spectrum obtained
by Cardy's condition to the open string spectrum
obtained in \cite{ooguri}
by direct quantization of the open string (see also 
\cite{Petropoulos:2001qu}).
We find exact agreement, which is strong support for
our approach.
Furthermore, our method works generally for all the branes.
and the open string spectrum obtained is always consistent.
(In fact, we find the open string spectrum to
be the same on all the branes.)
We believe this is convincing evidence that
our boundary states are exact.

We start by reviewing the closed string spectrum of  
\cite{mo}. We also quantize strings winding around the
closed timelike curve of $\SL$. We then review the 
D-branes, which were found by \cite{Stanciu:1999nx,
Figueroa-O'Farrill:2000ei,bachas} to wrap
conjugacy classes of  $\SL$. 

We then use this geometric description to construct the boundary state in
the large $k$ limit. This is done by requiring
that the overlap of D-branes with closed string modes
be restricted to the conjugacy class. This
allows us to write an expression for the boundary 
state which is valid in the large $k$ limit, and
more importantly, where the overlaps of boundary states
can be calculated. This expression is given at the
end of section 3.

Next, we compute the overlaps of these branes and check that
Cardy's condition is satisfied. This is a straightforward,
though technical, calculation. 
Finally, we close with a discussion of ${1\over k}$ corrections to the boundary
state, where we argue that the only effect of
corrections is to renormalize the overall constant
in front of the boundary state. 

A few comments are in order. First, our brane couples only to 
states in the continuous representations. This is in
contrast to the recent paper \cite{Hikida:2001yi}.
Secondly, only the unflowed representations appear in
the boundary state. Thirdly, the construction of the D-brane has many similar
aspects to the construction of D-branes in
Liouville theory \cite {armr2}. It would be interesting
to explore this further.

\section{ Closed Strings on $AdS_3$}
\label{sec1}
\subsection{Point Particles on $AdS_3$}

We review here the quantization of a point particle moving in 
an $AdS_3$ background. This serves to fix our notations, 
and introduce some formulae to be used later on.

In cylindrical (global) coordinates, the metric of $AdS_3$ is 

\beq
ds^2 = - \cosh^2\rho dt^2 + d\rho^2 + \sinh^2\rho d\phi^2
\eeq

We will often find it useful to replace the global coordinate system
$(\rho, \phi, t)$ by the coordinates $(\psi, \chi , t)$, with:

\beq
\label{coord}
\sinh\psi= \sinh \rho \sin \phi\qquad
\cosh\psi\sinh\chi=-\sinh\rho\cos\phi
\eeq

The $AdS_3$ metric in these coordinates is
\bea
ds^2=d\psi^2+\cosh^2\psi(-\cosh^2\chi dt^2 +d\chi^2)
\eea

The coordinate $\psi$ will be useful in describing branes wrapping twisted 
conjugacy classes, as they are located at constant values of $\psi$.

All modes of particles propagating in $AdS_3$ should fall into representations
 of the isometry group, namely $\SL_L\times \SL_R$. In general,
eigenfunctions of
the Laplacian on a group manifold can be written as:

\beq
D^j_{mm'}(x) = \langle jm|g(x)|jm'\rangle
\eeq
where $|jm \rangle$ is a basis for a representation of the  group, and
$g(x)$ is some parametrization of the group manifold. To make those functions
 explicit in our case we write  expressions for the generators of
 $\SL_L\times \SL_R$, as discussed by \cite{vijay}.

In global coordinates these generators are (see the Appendix for details)

\bea
\label{gens}
J^3&=& \frac{i}{2} \partial_u \nonumber \\
J^+ &=&\frac{i}{2} e^{-2iu}\left[ \coth 2\rho \partial_u-
\frac{1}{\sinh 2\rho } \partial_v + i \partial_\rho \right] \nonumber\\
J^-&=&
\frac{i}{2} e^{2iu}\left[ \coth 2\rho \partial_u-
\frac{1}{\sinh 2\rho }\partial_v- i \partial_\rho \right]
\eea
where 
\beq
u = \frac{1}{2} (t+\phi)~~~~~~~~~~~~ 
v=  \frac{1}{2} (t-\phi)
\eeq
The generators of the other $\SL$ algebra are obtained by exchanging
$u$ and $v$  in the above expressions.

We can now discuss the mode expansion for a massive scalar field, of mass $M$,
in $AdS_3$. The discussion follows the notation in \cite{vijay}.

The eigenvalue equation for the massive scalar field  is:

\beq
\Box \Phi = 
\partial_\rho^2 + 2 \frac{\cosh 2\rho }{\sinh 2 \rho } \partial_\rho
+ \frac{1}{\sinh^2 \rho }\partial_\phi^2 - \frac{1}{\cosh^2 \rho }
\partial_t^2 = M^2 \Phi
\eeq

The eigenvalues of the Laplacian are parametrized by $j$, 
which is defined through $ M^2= 4j(j-1)$.
%

The general mode can be characterized by 3 quantum numbers, which correspond
to 
$j$, and the two magnetic quantum numbers, $m$ and  $\bar{m}$,
in an $\SL \times \SL$ representation.

Explicitly, the eigenfunction of $J^2, J^3, \bar{J}^3$ is given by
(see \cite{absteg} for notation)
\beq
\Phi^j_{m\bar{m}} = C e^{-i\nu t} e^{il\phi} \cos^{2j}\mu 
\sin^l\mu  \ _2F_1(j+\frac{l+\nu}{2}, j+\frac{l-\nu}{2}, l+1;\sin^2 
\mu)
\eeq
where 
$\nu= m+\bar{m}, l = \bar{m}-m$,
and the coordinate $\mu$ is defined by $\tan\mu= \sinh\rho$.
$C$ is a normalization constant, which sets 
the overlap of $\Phi^j_{m\bar{m}}$ with itself to 1.

The states  of particles  on $AdS_3$ are given by (delta function)
normalizable states. This forces $j$ into one of two possible regions:

%

\begin{itemize}

\item{ $j> \frac{1}{2}$, $j$ real---discrete representations $D_j^\pm$ }

These are representations with highest (or lowest) weight state. They have
 real $j$
and a spectrum of magnetic quantum numbers $m$ which starts with
the highest (lowest) $m=j$, which is annihilated by $J^+$ (or $J^-$) and moves
down (up) a unit by repeated application of  $J^-$ (or $J^+$).
These representations are unitary for $j> \frac{1}
{2}$.

\item{$j= \frac{1+i s }{2} $ ($s$ real)---continuous representations $C^\alpha_j$ }

These do not have highest or lowest weight states, and therefore 
the spectrum of $m$ is unbounded from above and from below. The 
fractional part of $m$ is
preserved by an action of the creation and annihilation operators, 
and is denoted by $\alpha$. The spectrum is then $m= \alpha + k $ 
where $k$ is an arbitrary integer.

\end{itemize}

A complete basis of normalizable functions on $AdS_3$ is spanned by the 
representations $D_j^\pm \times D_j^\pm$ ($j>\frac{1}{2}$) and 
$C_j^\alpha \times C_j^\alpha$.
A state in one of those representations $|j,m\rangle \times |j,\bar{m}\rangle$
can be represented by the function $\Phi_{jm\bar{m}}$ defined above.

\subsection{Strings on $AdS_3$}

Quantization of strings on $AdS_3$ was performed in \cite{mo}, which we 
review here. The closed string spectrum found in \cite{mo}
is the starting point to building  boundary states for the D-branes propagating in $AdS_3$. 

For each
representation of the zero mode algebra $\SL\times \SL$ one can 
construct
 a module of the Kac-Moody algebra by a repeated application of oscillator
modes $J^a_{-n}, n>0, a= \pm, 3$. 
We denote such representations by $\hat{D}_j^\pm$ and
$\hat{C}_j^\alpha$. These representations were dubbed positive energy representations in \cite{mo}, as the spectrum of $L_0$ is bounded from below.

The no-ghost theorem of \cite{noghost,mo} states  that to ensure
unitarity (after imposing the Virasoro constraints) one must restrict
 the spectrum of the discrete representations
 $\frac{1}{2} < j < \frac{k-1}{2}$. There is no restriction on the 
spectrum of the continuous representations.

In \cite{mo}, a new set of representations of the KM algebra was constructed. 
These are obtained from the above representations by an application of the 
spectral
flow, defined as :
\beq
J_n^3 \rightarrow J_n^3 - \frac{k}{2}\omega \delta_{n,0} ~~~~
J_n^+ \rightarrow J_{n+\omega}^+ ~~~~
J_n^- \rightarrow J_{n-\omega}^- 
\eeq

This preserves the KM algebra, and was conjectured in \cite{mo} to be
a symmetry of the closed string spectrum. The new representations have 
$L_0$
unbounded from below, but satisfy a no-ghost theorem, proven in \cite{mo}.

The complete closed string
 spectrum then
 consists of the
representations allowed by the no-ghost theorem above, together with all
their
images under the spectral flow---the so-called flowed representations. We 
denote such representations by  $\hat{D}_j^{\pm, \omega}$ and
 $\hat{C}_j^{\alpha, \omega}$. Note that one can restrict attention to 
 $\hat{D}_j^{+, \omega}$ only, since $\hat{D}_j^{-, \omega}$ can 
be generated from it by spectral flow.

\subsection{Winding Strings on $AdS_3$}
\label{windsec}

For later use, we are interested in compactifying  time in the Euclidean 
version of the theory. This will be used as a regulator, as in \cite{mo}. The 
compactification introduces new sectors of closed strings, namely the winding
 strings. We quantize these strings, following \cite{mo}.

 Any string configuration is represented
by a map $g(\sigma,\tau)$ from the worldsheet to spacetime. As shown by 
\cite{mo}, the general such map which satisfies the equations of motion can 
be written as $g = 
g_+(x^+)g_-(x^-)$, where $x^\pm = \tau \pm \sigma$.

Starting with any such  configuration, one can obtain a   new one by  the 
transformation:
\be
g_+ \rightarrow e^{\frac{i}{2} \omega_R x^+ \sigma_2} g_+ ~~~~
g_- \rightarrow e^{\frac{i}{2} \omega_L x^- \sigma_2} g_+ 
\eeq

This is equivalent to
\bea
t &\rightarrow& t + \frac{1}{2}(\omega_L +\omega_R)\tau
+ \frac{1}{2}(\omega_R -\omega_L)\sigma \nonumber \\ 
\phi &\rightarrow& \phi + \frac{1}{2}(\omega_L +\omega_R)\sigma
+ \frac{1}{2}(\omega_R -\omega_L)\tau
\eea

This action with $\omega_l = \omega_R$ generates spectral flow.
If we are working on the universal cover of $AdS_3$, then the
time direction is noncompact. We are then forced to set $\omega_l = \omega_R$
so that fields are single valued.

On the single cover of $AdS_3$,
on the other hand, the time direction is compact with radius $2\pi$.
We can then take
$\omega_R = - \omega_L = w$.
The resulting configuration satisfies:
\beq
t(\sigma+2\pi) = t+ 2\pi  w
\eeq
i.e.
the string winds $w$ times around the time direction. 

More generally, if we compactify the time direction with the
identification $t \equiv t+ 4\pi R$, then taking $\omega_R = - \omega_L = 2wR$
generates a string winding $w$ times around the time direction.

The currents on the worldsheet are given by 
\bea
J^3 &=& k(\partial_+u + \cosh 2\rho  \partial_+v) ~~~~
\nonumber \\
J^\pm &=& k (\partial_+\rho \pm i \sinh 2\rho  \partial_+v) e^{\mp 2iu}
\nonumber \\
\bar{J^3} &=& k(\partial_-v + \cosh 2\rho  \partial_-u) ~~~~
\nonumber \\
\bar{J}^\pm &=& k (\partial_-\rho \pm i \sinh 2\rho  \partial_-u) e^{\mp 2iv}
\eea

The effect of creating winding $w$ described above is:
%
%
\bea
J^3 &\rightarrow&J^3+kwR ~~~~~~~\qquad
\bar{J^3} \rightarrow \bar{J^3}-kwR 
\nonumber\\
J^\pm&\rightarrow& e^{\mp2iRwx^+}J^\pm
~~~~\qquad
\bar{J^\pm} \rightarrow  e^{\pm2iRwx^-}\bar{J^\pm}
\eea

In terms of the modes, we have:
\bea
J^3_n &\rightarrow&
J^3_n +kwR \delta_{n,0} 
~~~~\qquad
\bar{J}^3_n \rightarrow\bar{J}^3_n -kwR \delta_{n,0} 
\nonumber \\
J^\pm_n &\rightarrow& J^\pm_{n\mp2wR}
~~~~~~\qquad\qquad
\bar{J}^\pm_n \rightarrow\bar{J}^\pm_{n\pm 2wR}
\eea

These formulae are similar to the ones in \cite{mo} for the spectral flow, 
except that the shifts on the two sides are opposite.
Thus
the operation here creates winding around the time direction, rather than
a long string winding. The stress tensor on the worldsheet is shifted to be:

\beq
\label{shift}
L_0 \rightarrow L_0 - 2wRJ_0^3 - kw^2R^2 ~~~~\quad\quad\qquad
\bar{L}_0  \rightarrow \bar{L}_0 + 2wR \bar{J}_0^3 -  kw^2R^2
\eeq

We will also need to consider winding strings for the
case of Euclidean $AdS_3$. This corresponds to
a Wick rotation of both $t$ and the worldsheet time $\tau$.
All formulae for this case are the same with the replacement
$R\rightarrow iR$.

\subsection{Geometry of $AdS_3$ D-branes}
\label{geom}
We turn now to the D-branes in $AdS_3$. First we
review the semiclassical, geometric, analysis performed in
\cite{bachas}. We concentrate on the physical 1-branes in Lorenzian
$AdS_3$\footnote{That is, worldvolumes which are 1+1 dimensional.}.
Those were found to wrap conjugacy classes which are generically $AdS_2$.
The conjugacy classes are topologically trivial, but
the wrapped branes are prevented from collapsing by a gauge field flux on
their worldvolume, as explained in \cite{bds}. Since the worldvolume is
non-compact, there is no quantization condition on the flux, and consequently
any conjugacy class is allowed (at the string tree level).

The 1-branes under consideration wrap twisted (twined)
conjugacy classes. These conjugacy classes are defined by having a constant 
value of
$tr(\sigma_1 g)$,
where $g$ is an element of $\SL$ in its parametrization by $2 \times 2$
matrices, and $\sigma_1$ is the Pauli matrix. The conjugation by the matrix
$\sigma_1$ generates the unique outer automorphism of $\SL$.
In cylindrical (global) coordinates these are the hypersurfaces:
\beq
\label{conj}
\sinh\psi=\sinh \rho \sin \phi  = C
\eeq
where $C$ is some constant.
Geometrically, these are static configuration of D-strings connecting 
antipodal  points of $AdS_3$. 

In addition to these 1-branes, one can construct branes wrapping regular 
conjugacy classes. Those are characterized by $tr(g)$ being a constant,
which translates to $\cosh\rho\cos t ={\rm const}$. 
Depending on the value of the constant, the regular conjugacy classes span 
$H_2$, the hyperbolic plane, or $dS_2$\ ---two dimensional deSitter 
space. While 
the former is appropriate to describing instantons in  $AdS_3$, the 
later were shown to correspond to unphysical D-branes, carrying 
hyper-critical electric field on their worldvolume. In the following we 
will
restrict our attention to the branes wrapping the twisted conjugacy classes only.

\section{Analysis in Conformal Field Theory}
\label{sec2}
\subsection{Ishibashi states}
\label{Ish}
We want to extend the semi-classical analysis to a full 
conformal field theory
treatment, by writing the boundary state which describes these D-branes.
The
condition in conformal field theory that the branes wrap 
the above conjugacy class is \cite{as,Felder:2000ka}
\beq
\label{cond}
(J^a_n + \bar{J}^a_{-n}) |B \rangle =0
\eeq
where $J^a, \bar{J^a}$ are the Kac-Moody currents on the worldsheet, and
$|B\rangle$ is the boundary state. 

Cognoscenti may be surprised at the above equation, which
is usually associated with regular conjugacy classes, not
twisted ones. 
This small peculiarity
follows from  our
definition of the currents $J^a, \bar{J^a}$. In our notation\footnote{This is
not 
just a matter of notation. In order to 
use the spectrum of representations in 
the
 closed string spectrum  as
stated in \cite{mo}, one has to adhere to their notation. Using different 
notation would entail translating their 
statements concerning the representation content as well.},
the condition (\ref{cond}) indeed implies (\ref{conj}).
  In particular, the branes wrapping the conjugacy classes defined in
(\ref{conj}) are static, therefore by energy conservation they 
 couple
(linearly) only to zero energy closed string states. 
 In other words, the boundary state for the twined conjugacy class
should 
satisfy $\del_t \ket{B}=0$, i.e. $(J^3+\bar{J}^3)\ket{B}=0
$
which is consistent with (\ref{cond}).

En passant, we note that the branes wrapping regular conjugacy classes
are characterized by the condition 
\bea
(J^3-\bar{J}^3)\ket{B}=(J^++\bar{J}^-)\ket{B}=(J^-+\bar{J}^+)\ket{B}=0
\eea

Following \cite{ishi, contra}, one might hope to find a solution to 
(\ref{cond})
for each module of the chiral algebra.
For our purposes it is sufficient to construct 
Ishibashi states based on continuous representations,  
$\hat{C}_j^{\alpha, \omega}$,
only. 
Indeed, the branes we discuss couple only to zero energy 
closed string states. In the   discrete representations  there 
are at most finitely 
many such states, insufficient number to construct a boundary state. 
It is only the continuous representations that have infinitely many 
zero energy 
states, which can then form a coherent state.  We conclude then that 
the boundary state we are seeking has an overlap with the 
representations $\hat{C}_j^{\alpha, \omega}$ only\footnote{This has to be 
contrasted with the recent discussion in 
\cite{Hikida:2001yi}. See also our remarks in 
the previous footnote.}.

In addition, the standard Ishibashi construction of a solution to (\ref{cond})
does not work for the discrete representations. The simplest way to see this 
is the following: the condition $(J_0^+  +\bar{J_0}^+) |B\rangle =0$
is usually satisfied by comparing the components of the boundary 
state $|B \rangle$
with magnetic quantum numbers $(m - 1, \bar{m})$ and $(m, \bar{m} - 1)$. 
 This generates a recursion relation whose solution is the set of 
coefficients of the Ishibashi state. 

However, here the spectrum of the magnetic quantum numbers is semi-infinite, 
so some of the states needed for the recursive cancellation are absent 
from the spectrum. This is then 
another argument to show that the discrete representations do not make an 
appearance in our boundary state. We are then only interested in Ishibashi states based on the continuous representations  $\hat{C}_j^{\alpha, \omega}$.

Furthermore, we are interested at first in the semiclassical, infinite $k$
limit, so we start by restricting  attention to the unflowed representations 
 $\hat{C}_j^{\alpha}$ only. 
The standard Ishibashi construction does work for the representations
 $\hat{C}_j^{\alpha}$. 
Suppose we are given a KM 
primary $|\Phi_j\rangle$ in the $\hat{C}_j^{\alpha}
\times \hat{C}_j^{\alpha}$ representation. By that we mean a state 
which is annihilated by all lowering operators $J_n^a, n>0$, but not
by   the zero modes (since 
$C_j^\alpha$ has no highest weight states). 
 Define\cite{contra,armr2}
\be
\label{contra}
|I^j\rangle = \sum_{I,J} M_{IJ}^{-1}J_{-I}
\bar{J}_{-J}|\Phi^j\rangle
\ee
Here $I,J$ are ordered strings of indices $(n_1, a_1)\cdots (n_r,a_r)$, and
\be
J_I=J_{n_1}^{a_1}\cdots J_{n_r}^{a_r}.
\ee

For later convenience we choose an ordering such that all zero modes 
act from the left. 
This sums over all the descendants in the KM module, with the normalization 
 defined as
\be
M_{IJ}=\langle \Phi^j| J_I\, J_{-J}|\Phi^j\rangle
\ee

$M_{IJ}$ is invertible for any KM module. (For degenerate
modules, one has to mod out by the null 
vectors).
It is easy to see that $|I^j\rangle$ satisfies (\ref{cond})  by
showing
that $(J_n+\bar{J}_{-n})|I^j\rangle$  is orthogonal to 
all states in the module based on
$|\Phi^j\rangle$.

We will also need to construct Ishibashi states when the
moding of the oscillators is shifted, as
in the winding string sector. The formulae are
very similar to the unshifted case, and are discussed
in e.g. \cite{armr1}.
\subsection{Overlaps of Ishibashi states-a difficulty}
\label{divdisc}

The boundary states $|I^j\rangle$ based on primaries in the 
 $\hat{C}_j^{\alpha}
\times \hat{C}_j^{\alpha}$ representations
are the building blocks of the desired boundary state. The physical 
boundary state is required
to satisfy Cardy's conditions \cite{cardy}, which guarantee the existence of 
open string quantization of the system. 

Here we point out  a difficulty in constructing a physical boundary state, 
which results from the fact that the representation of the zero mode algebra,
$C_j^\alpha$, is 
infinite dimensional. The overlaps between Ishibashi states are as usual 
the characters of the corresponding representations:
\beq
\langle I^j| q^{L_0 +\bar{L}_0 - \frac{c}{12}} |I^j\rangle = 
Tr_j(q^{2L_0- \frac{c}{12}})= \chi_j(q^2)
\eeq
where 
$\chi_j$ is the character in the representation based on
$\Phi^j$. This character diverges since  the character involves a sum over all 
magnetic quantum numbers $m$ (through descendants obtained by application of 
the zero mode
operators $J^\pm_0$), and $L_0$ is independent of $m$. This gives a
 divergence which needs regulating. We will find below    a way to 
represent this divergence which makes regularization straightforward.

A regularization which works for the discrete representation is replacing the
 character above by $tr_j(q^{2L_0- \frac{c}{12}}\, 
e^{2\pi i \theta J^3_0})$, where
$\theta$ is used as a regulator. This works well for the discrete 
representation, where it replaces the divergence by a well behaved modular 
functioni \cite{Gawedzki:1991yu}. For the continuous representations, however, this regularization  
 yields a result proportional to  $\delta(\theta)$, which is an awkward object 
to manipulate.

We emphasize that the difficulty has to do with the zero modes, and not with 
the string oscillator modes. It will be already present when quantizing a 
particle on $AdS_3$.
In an attempt to separate the stringy aspects from the 
infrared
 aspects of 
the boundary states, we define the following  coherent state:
\beq
\label{ishdef}
|I^j_{m\bar{m}}\rangle_ = \widetilde{\sum_{I,J}} M_{IJ}^{-1}J_{-I}
\bar{J}_{-J}|\Phi^j_{m\bar{m}}\rangle
\eeq

Here $|\Phi^j_{m\bar{m}}\rangle$ is annihilated by all oscillator modes except 
for the zero modes (as above), and has magnetic quantum numbers $m$ 
and $\bar{m}$. The sum
$\widetilde{\sum}$ over descendants here is defined to 
exclude any action by the 
zero
 modes $J^\pm_0$. It is clear by definition that:
\beq
\label{ish}
|I^j \rangle = \sum_{m} |I^j_{m,-m}\rangle\
\eeq

The overlap of these coherent states is finite. 
It is possible to exhibit this overlap  as a product of a 
contribution from the primaries and a contribution
of the stringy oscillators:
\beq
\label{CtoB}
\langle I^j_{m\bar{m}}| q^{L_0 +\bar{L}_0- \frac{c}{12}} 
\, e^{\pi i \theta (J^3_0 - \bar{J^3_0})}|I^j_{m\bar{m}} \rangle =
\\
{\langle \Phi^j_{m\bar{m}}| q^{L_0 +\bar{L}_0- \frac{c}{12}}
\, e^{\pi i \theta (J^3_0 - \bar{J^3_0})}|\Phi^j_{m\bar{m}} \rangle\over
 \prod_{n=1}^\infty (1-q^{2n})(1-q^{2n} e^{2\pi i  \theta})(1-q^{2n} 
e^{-2\pi i  \theta}) }
\eeq

The overlap of the primaries written above is 
\bea
\langle \Phi^j_{m\bar{m}}| q^{L_0 +\bar{L}_0- \frac{c}{12}}
\, e^{\pi i \theta (J^3_0 - \bar{J^3_0})}|\Phi^j_{m\bar{m}} \rangle
=q^{\left(-\frac{2j(j-1)}{k-2}-\frac{k}{4(k-2)}\right)}e^{ \pi i 
 \theta (m-\bar{m})}
\eea

\subsection{From geometry to CFT}

The full CFT description requires us to find a linear
combination of Ishibashi states, which satisfies Cardy's
condition. This is rendered somewhat difficult
because of the divergences discussed in subsection 
(\ref{divdisc}).
We will see now how the 
geometry of the
$AdS_3$ branes (discussed in subsection (\ref{geom})) helps us to 
understand the boundary state description of
the D-branes.

The connection between the boundary state and the D-brane
geometry is well-known in the
study of the $SU(2)$ WZW model. This is most clearly
explained in Appendix B of \cite{seib}; we
review their analysis here.

Define a graviton wavepacket $\ket{x}$ localised at
a point $x$ on the group manifold, i.e.
\bea
\langle x\ket{\Phi}=\Phi(x)
\eea

In the semiclassical limit, the D-brane is
localized at the conjugacy class $\psi=\psi_0$.
Hence the overlap of the D-brane with the graviton
wavepacket described above should also be localized
on the conjugacy class i.e.
\bea
\label{limeq1}
\lim_{k\rightarrow \infty}\  \langle x\ket{B}={f(\psi_0)\over \cosh\psi_0}\d(\psi-\psi_0)
\eea
where $f(\psi_0)$ is a function proportional to the tension
of the brane.

In \cite{Felder:2000ka,seib}, it was shown that a formula of the above form
indeed holds for the branes in the $SU(2)$
WZW model i.e. in the large $k$ limit,
the boundary states are found
to
be located on conjugacy classes. 

We will reverse the procedure for the $\SL$ case; given
the geometry of the brane, we will obtain
information about the boundary state.

The boundary state is in general a linear superposition of 
Ishibashi states 
\bea
\ket{B}=\sum_{jm\bar{m}} c^j_{m\bar{m}}\ket{I^j_{m\bar{m}}}
\eea
where $\ket{I^j_{m\bar{m}}}$ is the coherent  state 
based  
on the primary $\Phi^j_{m\bar{m}}$ as in (\ref{ishdef}).
The $ c^j_{m\bar{m}}$ are constants which are
the main data required for specifying the boundary state.
We shall determine 
the coefficients
 $ c^j_{m\bar{m}}$ (to leading order in ${1\over k}$) by
using  equation  (\ref{limeq1})\footnote{In the SU(2) case, the
coefficients were already known from the work of \cite{cardy,ishi}.}.

Combining the various equations, we see that
the required equation for the
coefficients is
\bea
\label{limeq2}
\sum_{jm\bar{m}} c^j_{m\bar{m}}(\psi_0) \Phi^j_{m\bar{m}}(\psi,\chi,t)=
{f(\psi_0)\over \cosh\psi_0}\d(\psi-\psi_0)
\eea
where we have exhibited the dependence of the coefficients  $c^j_{m\bar{m}}$ 
on the conjugacy class, parametrized by its location $\psi_0$.

Furthermore, the construction of the Ishibashi state
tells us that all $\Phi^j_{m\bar{m}}$ with 
$m +\bar{m}=0$ contribute equally, i.e.
\beq
 c^j_{m\bar{m}}=c^j\delta_{m+\bar{m},0}
\eeq

So the boundary state can be written
\beq
\ket{B}=\sum_j c^j\sum_m\ket{I^j_{m,-m}}
\eeq
and  equation (\ref{limeq2}) simplifies to
\bea
\sum_j c^j(\psi_0) \sum_m \Phi^j_{m,-m}(\psi,\chi,t)={f(\psi_0)\over \cosh\psi_0}\d(\psi-\psi_0)
\eea

Our task is now to invert this equation and obtain the 
coefficients  $c^j(\psi_0)$.
Note that the overlap of two such boundary states is still divergent, 
since it includes the following  factor, resulting 
from the overlap of primaries:
\bea
\langle B|q^{L_0}\ket{B} \propto \sum_j |c^j|^2\sum_m \langle{\Phi^j_{m,-m}}
|q^{L_0}\ket{\Phi^j_{m,-m}}
=
\sum_j |c^j|^2\sum_m q^{-j(j-1)/(k-2)}
\eea
Since we have an infinite sum over $m$, this expression diverges.

This divergence is not regulated by separating the branes e.g. if
we separate the branes by rotating one of them through
an angle $\theta$, the overlap  contains the factor 
\bea
\langle B|q^{L_0}e^{i\theta(J^3_0-\bar{J^3_0})}\ket{B}
\propto \sum_j |c^j|^2\sum_m q^{-j(j-1)/(k-2)}e^{2im\theta}
=\sum_j |c^j|^2q^{-j(j-1)/(k-2)}\d(\theta)
\eea

What we have found is exactly the divergence discussed
earlier, where we found that the characters of the 
continuous representations were ill defined. However, we presented here the 
divergence in a form amenable to regularization, which we now perform.

\subsection{The correct approach}

As we have seen, our equation (\ref{limeq2}) has reduced to
\bea
\sum_j c^j(\psi_0) \sum_m \Phi^j_{m,-m}(\psi,\chi,t)={f(\psi_0)\over \cosh\psi_0}\d(\psi-\psi_0)
\eea

So instead of considering each $\Phi^j_{m,-m}$, we see that
the only relevant combination we need to consider
is
\bea
\label{phidef}
\Phi^j=\sum_m \Phi^j_{m,-m}
\eea

Surprisingly, we see that if we use $\Phi^j$, all our problems
can be resolved!

First, to determine $\Phi^j$ as a function of $(\psi, \chi, t)$, note that  
from the definition above, we have
$(J^a+\bar{J}^a)\Phi^j=0$. 

Using (\ref{gens}),  
\bea
(J^3+\bar{J}^3)\Phi^j=0\quad\Rightarrow\quad \del_t\Phi^j=0.
\eea

The equation  $(J^++\bar{J}^+)\Phi^j=0$
can then be written
\bea
e^{-2iu}\left(
{\cosh2\r\over \sinh2\r}\del_u-{1\over \sinh2\r}\del_v+{i\over2}\del_\r\right)
\Phi^j(\psi,\chi)~~~~~~~~~~~~~~~~~~~~~~~~~~~~~~~~
\nonumber
\\
~~~~~~~~~~~~~~~~~~~~~~~~~~~~~~~~~~+e^{-2iv}\left(
{\cosh2\r\over \sinh2\r}\del_v-{1\over \sinh2\r}\del_u+{i\over2}\del_\r\right)
\Phi^j(\psi,\chi)=0
\eea
which reduces to
\bea
\del_\chi \Phi^j (\psi,\chi)=0
\eea

Hence we find that $\Phi^j$ is a function of $\psi$ alone.

Since $\Phi^j$ was defined as a linear combination
of 
 $\Phi^j_{m,-m}$, which are all eigenfunctions of the
Casimir with eigenvalue $4j(j-1)$, the same holds
true for $\Phi^j$ i.e.
\bea
\Box \Phi^j=4j(j-1)\Phi^j
\eea

Since $\Phi^j$ is a function of $\psi$ alone, this equation becomes
\bea
\del_\psi^2\Phi^j+{2\sinh\psi\over\cosh\psi}\del_\psi \Phi^j=4j(j-1)\Phi^j
\eea

The two independent solutions to this differential equation are
\footnote{The $\Phi^j$ here are related to the definition in
(\ref{phidef}) by an overall constant.}
\bea
\Phi^j= {e^{\pm(2j-1)\psi}\over \cosh\psi}
\eea 

Note that if $(2j-1)>0$ then $\Phi^j$ diverges either at
$\psi=\infty$ or $\psi=-\infty$. If we want 
$\Phi^j$ to be normalizable,
we must take $(2j-1)=is$ with $s$ real. Comparing
with the closed string spectrum in section \ref{sec1}, this implies that 
$\Phi^j$ is a combination of states in the continuous 
representations alone, and discrete states do not
have any overlap
with the D-brane. 
This is consistent with the analysis in section  \ref{sec2}.

Now that we know $\Phi^j$, we return to equation (\ref{limeq2}),
\bea
\sum_j c^j(\psi_0) \Phi^j(\psi)={f(\psi_0)\over \cosh\psi_0}\d(\psi-\psi_0)
\eea
which we can write as
\bea
\sum_s c^s(\psi_0) 
{ e^{is\psi}\over \cosh\psi}
={f(\psi_0)\over \cosh\psi_0}\d(\psi-\psi_0)
\eea

The obvious solution is that
$c^s(\psi_0)$ is proportional to  $e^{-is\psi_0}$.
The sum over $s$ will then yield a delta function in $(\psi-\psi_0)$.
The function $f(\psi_0)$ cannot be determined
at this level. We will later use Cardy's condition
to determine it.

Thus we  have  found the boundary state in the semiclassical
approximation to be

\beq
\label{Bdef}
|B \rangle = f(\psi_0) \sum_s e^{-is\psi_0} |I^s \rangle
\eeq
where $I_s$ is the Ishibashi state (as defined
in (\ref{ishdef})) based on the primary $|\Phi^s \rangle$,  satisfying
$\langle x |\Phi^s \rangle = \frac{e^{is\psi}}{\cosh\psi}$.

We now turn into a calculation of the overlap of different boundary states. 
This exhibits the regularization of the above discussed divergence, and 
will serve as the basis of comparison to the open string sector, and to the 
discussion of Cardy's condition that follows. 

\section{Cardy condition 1---Overlaps of branes}

We will compute four different overlaps in this section.

a) The overlap of two branes, one located at $\psi=\psi_0$,
and the other at $\psi=\tilde{\psi_0}$.

b) The overlap  after one of the branes in (a) is rotated through
an angle $\phi_0$.

c) The overlap of the branes in (a), when they exchange winding modes.

d) The overlap of the branes in (b), when they exchange winding modes. 

\subsection{Notation}
 We will be considering two D-branes 
 labeled by $\psi_0$ and $\tilde{\psi_0}$, so that
\beq
|B \rangle = f(\psi_0)  \sum_s e^{-is\psi_0} |I^s \rangle ~~~~\qquad
|\tilde{B} \rangle = f(\tilde{\psi}_0)  \sum_s e^{-is\tilde{\psi}_0} |I^s 
\rangle
\eeq

 Overlaps
of boundary states are defined through
$\langle \tilde{B}| q^{L_0+\bar{L}_0-{c\over 12}}\ket{B}$, which represents 
the annulus diagram as calculated in the closed string sector. 
When computing
the overlap, it will prove convenient at intermediate stages to
restrict to the low energy limit,
ignoring stringy oscillators. For this purpose we define:
\beq
|C \rangle =  \sum_s e^{-is\psi_0} |\Phi^s \rangle ~~~~\qquad
|\tilde{C} \rangle =  \sum_s e^{-is\tilde{\psi}_0} |\Phi^s \rangle
\eeq

This only has overlap with primaries. We can then  relate overlaps of 
$|B\rangle$
and $|C \rangle$ by using (\ref{CtoB}).

We will also need the formula:
\beq
\left(L_0 + \bar{L}_0 -\frac{c}{12}\right)  |\Phi^s \rangle
= \left(\frac{-2j(j-1)}{k-2} - \frac{k}{4(k-2)}\right)  |\Phi^s \rangle =
\left(\frac{s^2}{2(k-2)} -\frac{1}{4}\right) |\Phi^s \rangle
\eeq

Also we define $q=e^{i\pi \tau}$.

\subsection{Overlaps of parallel branes}

The overlap of the branes described above is
\bea
\langle\tilde{C}|q^{L_0 + \bar{L}_0 -\frac{c}{12}}
|C \rangle 
= q^{-\frac{1}{4}} \sqrt{\frac{k-2}{2i\pi^2\tau}}
\exp\left({ (\tilde{\psi_0}-
\psi_0)^2 (k-2)\over  2i\pi \tau}\right)
\eea

Hence using equation (\ref{CtoB})

\bea
\langle\tilde{B} | q^{L_0 + \bar{L}_0 -\frac{c}{12}}
|B \rangle =
f(\psi_0) f(\tilde{\psi_0})
 \sqrt{\frac{k-2}{2i\pi^2\tau}}
\exp\left({ (\tilde{\psi_0}-
\psi_0)^2 (k-2)\over  2i\pi \tau}\right)
\frac{2\pi}{\theta_1'(0,\tau)}
\eea

\subsection{Overlaps of rotated branes}

Consider now the case when the two branes are rotated with respect to each
other in the $\phi$ direction. If a brane is located at $\sinh{\psi}
= \sinh\rho \sin \phi = \sinh \psi_0 $, then after rotation 
it will be located at $
\sinh\rho \sin (\phi- \phi_0) = \sinh\psi_0
$ i.e. 

\beq
\sinh\psi\cos\phi_0 + \cosh\psi\sinh\chi\sin\phi_0 = \sinh\psi_0
\eeq
(cf. coordinate system at equation (\ref{coord}).)

We will compute the overlap between the above described brane and a brane located at $\sinh\psi = \sinh\tilde{\psi_0}$.

From a CFT point of view we are computing:
\beq
\langle\tilde{B} |q^{L_0 + \bar{L}_0 -\frac{c}{12}}e^{-i\phi_0 (J_0^3-
\bar{J_0^3})}
|B \rangle
\eeq
Here $J_0^3$ and
$\bar{J_0^3}$ are the zero modes on the worldsheet of the currents $J^3$ and
$\bar{J^3}$. 

We compute first
 \bea
\langle\tilde{C} |q^{L_0 + \bar{L}_0 -
\frac{c}{12}}
e^{-i\phi_0 (J_0^3- \bar{J}_0^3)}
|C \rangle =
\int dx \ \langle\tilde{C}|q^{L_0 + \bar{L}_0 -\frac{c}{12} }
|x \rangle
\langle x|
e^{-i\phi_0 (J_0^3- \bar{J_0^3})}
|C \rangle
\eea

Here $|x\rangle$ is a complete basis of position eigenstates
 and $\int dx |x\rangle \langle x| =1$.

By definition of the rotated brane:
\beq
\langle x|e^{-i\phi_0 (J_0^3-
\bar{J_0^3})}
|C \rangle = \delta(\sinh\psi\cos\phi_0 + \cosh\psi\sinh\chi\sin\phi_0 - \sinh\psi_0)
\eeq

From the boundary state description of $\tilde{C}$
\beq
\langle\tilde{C} |q^{L_0 + \bar{L}_0 -\frac{c}{12}}
|x \rangle = \int {ds\over 2\pi} q^{-\frac{1}{4}} q^{(\frac{s^2}{2(k-2)})}
e^{i\tilde{\psi}_0s}{e^{-i\tilde{\psi}s}\over \cosh\psi}
\eeq

The overlap  
$\langle\tilde{C} |q^{L_0 + \bar{L}_0 -\frac{c}{12}}e^{-i\phi_0 (J_0^3-\bar{J}_0^3)}\ket{C} $ is then 
\bea
k^{3/2}\int dtd\psi d\chi \cosh^2\psi\cosh\chi ~~~~~~~~~~~~~~~~~~~~~~~~~~~~~~~~~~~~~~~~~~~~~~~~~~~~~~~~~~~~
\nonumber \\
 \times \ \d(\sinh\psi\cos\phi_0+
\cosh\psi\sinh\chi\sin\phi_0-\sinh\psi_0) ~~~~~~~~~~~~~
\nonumber \\
~~~~~~~~~~~~~~~~~~~~~~~~~~~~~~~~~~~~~~~~~~~~~~~~~~~~\times
\int {ds\over 2\pi} q^{-\frac{1}{4}} q^{(\frac{s^2}{2(k-2)})}
e^{i\tilde{\psi}_0s}{e^{-i\tilde{\psi}s}\over \cosh\psi}
\nonumber \\
=
 {k^{3/2}R_t\over |\sin\phi_0|}\int {ds}  q^{-\frac{1}{4}} q^{(\frac{s^2}{2(k-2)})}
\d(s)~~~~~~~~~~~~~~~~~~~~~~~~~~~~~~~~~~~~~~~~~~~~~~~~~~~
\nonumber\\
={k^{3/2}R_t\over|\sin\phi_0|} q^{-\frac{1}{4}}~~~~~~~~~~~~~~~~~~
\eea
where we have assumed the time direction
to be compact with radius $R_t$.

So, using (\ref{CtoB}), the full overlap is
\bea
 \langle\tilde{B} |q^{L_0 + \bar{L}_0 -\frac{c}{12}}e^{-i\phi_0 (J_0^3-
\bar{J_0^3})}\ket{B}
={2f(\psi_0)f(\tilde{\psi}_0) k^{3/2}R_t\over 
|\theta_1\left( {\phi_0\over \pi},\tau\right)|}
\eea

Note that only the $s=0$ term contributes.

\subsection{Overlaps of parallel branes-winding sectors}

Cardy's condition requires us to interpret an overlap of branes as
a modular transformation of an open string partition function. In
usual cases, (e.g. $SU(2)$ WZW models), this requires the
modular transformation to be a linear combination of
open string characters with positive integer coefficients.

Unfortunately, the $\SL$ model has a continuum of states (the long string
states \cite{Giveon:1998ns}) and it is difficult to extract the
discrete states out of this continuum.
A method for doing this was proposed in \cite{mo}. For our
purposes, duplication of
this method requires us to compactify
the time direction, and
compute overlaps of branes as
a function of the radius of the time
direction.

The nontrivial dependence on the radius occurs because now there are new
states in the theory with which the boundary state has overlap, viz.
strings winding around the time direction. Exchange of such
winding modes leads to new terms in the overlap of
the boundary states.

In general, the boundary state can now be written
\bea
\ket{B}=\sum_w \ket{B_w}
\eea
where $\ket{B_w}$ is the part of the boundary state
$\ket{B}$ having overlap with strings of winding number $w$.

The overlap is now a sum of terms
\bea
\label{totov}
\langle B|q^{L_0 + \bar{L}_0 -\frac{c}{12}}\ket{B}=\sum_w\langle B_w |q^{L_0 + \bar{L}_0 -\frac{c}{12}}\ket{B_w}
\eea

To compute this overlap, we note that the single cover
of $\SL$ has a compact timelike direction. There is a group transformation that
transforms a string winding $w$ times around this compact timelike
directon to an unwound string (as discussed in  (\ref{windsec})). This is very similar
to the transformation that produces spectral flow. Since
spectral flow is believed to be an exact symmetry, it
is plausible that is other transformation
is also an exact symmetry, and we conjecture that it is.
This transformation should leave the boundary state unchanged, and
hence we should have $\ket{B_w}\rightarrow \ket{B_0}$.

From equation (\ref{shift}), the transformation also acts on the energy as a shift
\bea
(L_0+\bar{L}_0)\rightarrow (L_0+\bar{L}_0)'=(L_0+\bar{L}_0)
-(2wR)(J_0^3-\bar{J}_0^3)-2kw^2R^2
\eea

This together implies
\bea
\langle \tilde{B}_w |q^{L_0 + \bar{L}_0 -\frac{c}{12}}\ket{B_w}
=q^{-2kw^2R^2}\langle \tilde{B}_0 |q^{L_0 + \bar{L}_0 -\frac{c}{12}}
e^{-i\phi_0(J_0^3-\bar{J}_0^3)}\ket{B_0}
\eea
where $\phi_0=2\pi wR\tau$.

This overlap was already calculated in the previous subsection. We find
\bea
\langle \tilde{B}_w |q^{L_0 + \bar{L}_0 -\frac{c}{12}}\ket{B_w}
=q^{-2kw^2R^2}(8\pi k^{3/2}R){f(\psi_0)f(\tilde{\psi}_0)\over 
|\theta_1\left(
2wR\tau,\tau
\right)|}
\eea

Since we get the same contribution from $w$ and
$-w$, we can restrict the sum to run over the positive integers
only. Then the overlap is
\bea
\sum_{w=1}^{\infty}q^{-2kw^2R^2}(16\pi k^{3/2}R){f(\psi_0)f(\tilde{\psi}_0)\over 
|\theta_1\left(
2wR\tau,\tau
\right)|}
\eea

This is the overlap of branes in Lorentzian $AdS_3$. In Euclidean
$AdS_3$, the overlap is obtained by replacing $R$ by $iR$.

\subsection{Overlaps of rotated branes (winding sector)}

Finally, one can compute the overlaps of relatively rotated branes in the winding
sector. This is a simple combination of the two previous subsections. The result
is:

\beq
\langle \tilde{B}|  q^{L_0 +\bar{L}_0 -\frac{c}{12}}e^{-i\phi_0(J_0^3- 
\bar{J_0^3}) } |B\rangle =
\sum_{w=1}^{\infty} q^{-2kw^2R^2}
 f(\psi_0)f(\tilde{\psi}_0)\frac{16\pi k^{3/2} R}
{|\theta_1(2wR\tau+{\phi_0\over \pi}, \tau)|}
\eeq

\section{Some digressions}
\subsection{The vanishing divergences}
We have obtained a finite answer for the overlap of the boundary states. 
Where did the divergences in the characters disappear?

It turns out that these divergences are hidden in the infinite volume of 
the D-branes. We found the D-brane is a source for $\Phi^j$, which is a constant
 over the conjugacy class. The overlap $\langle \Phi^j  | \Phi^j\rangle$ is 
therefore 
proportional to the volume squared of the conjugacy class, and hence divergent.

Once the divergence is presented this way, it is easy to regulate. One 
normalizes the boundary state by a factor $\frac{1}{\sqrt{V}}$, where $V$ 
is the volume. The resulting overlap then scales like the volume, which 
is the correct scaling for the open string partition function. 
On the other hand the overlap of the rotated branes  is now finite, as it 
should be since the open strings are now fixed at the intersection point.
Thus the delta function found earlier for the rotated branes is now 
regulated as well.

Note that the field $\Phi^j$ we found is strictly 
speaking not a normalizable mode, 
since its overlap 
diverges. This is because this field is an infinite sum of normalizable 
modes. 
(Hence the boundary state only couples to normalizable modes.)

\subsection{Relation to the $SU(2)$ Boundary States}

The $SL(2,R)$ group manifold is related to the $SU(2)$ group manifold by 
analytic continuation. We might expect the boundary states to be related 
this way. This is not the case.  For example, there is an analogue of the 
field $\Phi^j$
in the  $SU(2)$ case, which is equal to 
$\frac{\sin((2j+1)\tilde{\psi})}{\sin{\tilde{\psi}}}$. On the other hand,
in the $\SL$ case,  we had
two independent solutions for each $j$, which were 
$\Phi^j = \frac{e^{\pm(2j-1)\psi}}{\sinh\psi}$.

The reason for the difference is that in the  $SL(2,R)$ case there are two 
continuous representations, with $j = \frac{1}{2} \pm is$, having the same 
Casimir, whereas in the $SU(2)$ case there is only one representation for 
each Casimir.

This appears to be a generic difference between compact and non-compact WZW 
models. One important effect of this is that we can completely localize the 
$AdS_3$ branes, whereas the  $SU(2)$ branes are not fully localized on the 
conjugacy classes for finite $k$.

Furthermore, if the $\SL$ theory had contained finite
dimensional representations, then boudary states built
on these representations would have been very similar to
$SU(2)$ boundary states \cite{Giveon:2001uq}.

\subsection{Spectral Flow}

So far we have considered only the unflowed representations. We turn now to the
flowed representations---those constructed by acting with spectral flow on
the positive energy representations.  We argue here that flowed representations
do not contribute to the boundary state.

Intuitively, the flowed representations correspond to states of long strings
winding near the boundary of $AdS_3$. 
The D-branes do not wind near the boundary, so classical long string 
configurations cannot be absorbed by the
branes.

Nevertheless, the long string winding number is not conserved when the string
moves into the bulk, so there is some possibility of absorption by the brane.
This seems implausible to us, since locally the brane we constructed looks like
a flat brane in flat space. The overlap with the unflowed representations
accounts for all the states we need to reconstruct this flat space limit.
An overlap with the flowed representation is then possible only if there is
some subtlety in the flat space (large $k$) limit.

The above heuristic arguments are only semi-classical. The most convincing
argument is our ability to construct an exact Cardy state without any
recourse to flowed representations. This argument relies on an exact CFT
analysis and therefore is not limited to the semiclassical regime.

Two more comments are in order. First, the restriction to unflowed
representations does not truncate the spectrum (as it would for discrete
represenatations), and therefore there is no problem with reproducing
the  flat space boundary state. Secondly, the open string spectrum can and
does carry long string
winding number. Indeed, the modular transform of the Cardy state we construct
reproduces the open string result of \cite{ooguri}.

\subsection{Internal CFT and ghosts}

To construct a complete bosonic critical string theory containing an $AdS_3$
factor we need to add an
extra matter CFT as well as $b,c$ ghost system. An example of an
extra matter CFT is $S^3 \times T^{20}$.

The boundary states and their overlaps  factorize into contributions
from the 3 separate CFT factors. We need certain properties about the modular
transformation of the overlaps of the matter CFT and the ghost sector.

Under $\tau \rightarrow -\frac{1}{\tilde{\tau}}$ we have, for the ghost sector
(up to overall constants)
\beq
\label{modgh}
 _{gh}\langle B| q^{L_0 + \bar{L}_0-\frac{c}{12}} |B\rangle_{gh}
\quad
\rightarrow
\quad
{\tilde{q}^{1/6}\over \tilde{\tau}}
\prod_n [1-\tilde{q}^{2n}]^2
\eeq

Secondly, we will assume that we have already found a Cardy state in
the internal CFT, i.e. the modular transform of that part
of the boundary state yields a well defined open string 
partition function.

The total central charge of the internal CFT on the
open string is $c_{int}= 26-c_{AdS_3} =26-
\frac{3k}{k-2}$. The modular transform of the 
part of the boundary state depending on the internal CFT $M$ 
is then of the form (up to overall constants)
\beq
\label{modM}
_M\langle B| q^{L_0 + \bar{L}_0-\frac{c}{12}} |B\rangle_M
\quad
\rightarrow
\quad
\tilde{q}^{-\frac{c_{int}}{12}}\sum_h D(h) \tilde{q}^{2h}
\eeq
where $h$ is the weight in the open string conformal field
theory, and $D(h)$ is the degeneracy at weight $h$.

\section{Cardy condition 2---Open string partition function}
\subsection{$\psi_0=0$}
We now modular transform the brane overlaps to obtain the open string partition
function. We start by looking at the 
overlap in the winding sector of two branes,
both located at $\psi_0 =0$. We can compare 
this to the computation in Appendix B
of \cite{ooguri}. 

The open string partition function, calculated 
in Euclidean $AdS_3$  is  \cite{ooguri}
\bea
Z(\beta) = \frac{\beta\sqrt{k-2}}{4 \pi\sqrt{2}} 
\sum_{m=1}^\infty
{1\over t^{3/2}}
e^{2\pi t (1- {1\over 4(k-2)})} 
\sum_h
D(h) e^{-2\pi t h} ~~~~~~~~~~~~~~~~~~~~~~~~~~~~~~
\nonumber
\\
~~~~~~~~~~~~~~~~~~~~~~~~~~~~~~\times\ \frac{e^{-(k-2)\frac{\beta^2 m^2}{8\pi t}}}
{\sinh(\frac{\beta m}{2})} 
\prod_{n=1}^\infty \frac{|1-e^{-2\pi t n }|}
{|(1-e^{-2\pi t n+\beta m })(1-e^{-2\pi t n-\beta m })|}
\eea

On the other hand, we found that the overlap of branes in Euclidean $AdS_3$ 
to be
\beq
16\pi ik^{3/2} R f(\psi_0)f(\tilde{\psi_0})\sum_{w=1}^{\infty} \frac{
q^{2k w^2 R^2}}{|\theta_1(2i w R \tau
, \tau)|}
\eeq

Defining
$\b=4\pi R, \tilde{\tau}=it,\tilde{q}=e^{i\pi\tilde{\tau}}$, 
the modular transform 
$ (\tau = -\frac{1}{\tilde{\tau}})$ of this expression is 

\beq
Z= 4f^2(0)k^{3/2} \sum_{w=1}^{\infty}\frac{e^{-(k-2)\frac{\beta^2 w^2}{8\pi t}}e^{\frac{\pi t }{4}}
( \beta) }{\sqrt{t} \sinh(\frac{\beta w }{2}) 
|\prod_{n=1}^\infty (1-\tilde{q}^{2n}) (1-\tilde{q}^{2n}e^{-\beta w})
 (1-\tilde{q}^{2n}e^{\beta w})|}
\eeq

The full overlap is then found by multiplying the
above expression by the factors coming from the 
ghost sector and the internal matter sector (\ref{modgh}, \ref{modM}).
We see that we get exact agreement with the result from
the open string sector
(once $f(0)$ is chosen to match the overall constants).

\subsection{$\psi_0\neq 0$}
Now we look at the overlap between two branes labeled by $\psi_0$. This 
overlap is identical to the one of branes located at $\psi_0 =0 $, except 
for an additional factor $\frac{f^2(\psi_0)}{f^2(0)}$.
On the open string side the partition function picks up the same factor. This 
means that the degeneracies of open string states is multiplied by that
factor. Since $\psi_0$ is a continuous variable, and the degeneracies are 
integral, we conclude that  $\frac{f^2(\psi_0)}{f^2(0)} =1$
(to all orders in ${1\over k}$).

This also means that the open string spectrum on all branes is identical. We 
discuss this further below.

We can hence finally write the complete boundary state as
\bea
\ket{B}_{tot}=T\ \ket{B}\ket{B}_M\ket{B}_{gh}
\eea
where $T$ is an overall $k$ dependent factor, which represents the
tension of 
the boundary state (for flat space, the value of $T$ was found in
\cite{DiVecchia:1997pr}).
$\ket{B}$ was defined in (\ref{Bdef}), and $\ket{B}_M, \ket{B}_{gh}$
are the boundary states for the internal CFT and the ghost sector.

\subsection{Rotated branes}
The overlaps of rotated branes yield an interesting structure. The modular 
transform in the zero winding sector is (up to an overall constant)

\beq
\frac{\tilde{q}^{-\frac{\phi_0^2}{\pi^2}}}{\sqrt{\tilde{\tau}}\ 
\theta_1(-\frac{\phi_0 
\tilde{\tau}}{\pi},
\tilde{\tau})}
\eeq

This is very similar to the overlap in the closed string sector with winding.
  This suggests that the open strings are now described by a shifted stress  
tensor:

\beq
L_0 \rightarrow L_0 +\frac{\phi_0}{\pi}(J^3-\bar{J}^3) - \frac{\phi_0^2}{\pi^2}
\eeq

It would be very interesting to see this shift directly in the open string
side. The quantization of the open string is subtle, since one has to define 
the WZ term on worldsheet with boundaries.

\section{Discussion and conclusion}

Our calculations so far are valid only in the large $k$ limit, where we have 
a geometrical description. Can we go beyond this limit? We argue now that 
it is plausible that the boundary state described here is exact for all $k$.

At first  our result, that 
the open string spectrum is
 the same on all the branes, might seem surprising.  One might expect 
smaller branes to have a different spectrum than  the larger branes.

However, this can be anticipated  based on 
 \cite{bachas}. It turns out that the 
open string  
metric,
in the sense of \cite{sw},  is the same on all branes
once the electric field is taken into account. This provides  some evidence 
that  
the open string spectrum is indeed universal.

Secondly, in \cite{ooguri} it was argued that the 
density of 
states of long strings with odd spectral flow  was different for 
different branes. This was argued based on semiclassical quantization
of the open string, which is valid to leading order in $k$. At large
$k$, our results are in agreement with \cite{ooguri}. 
Indeed, the 
above mentioned difference in densities is subleading in $\frac{1}{k}$.

Moving to finite $k$, we argue that the boundary state is exact (upto
an overall $k$ dependent factor).
As we have seen,  the boundary states satisfy 
 Cardy's condition for any $k$.
Indeed, using our boundary states localized at any value of $\psi_0$,
we reproduce the result of \cite{ooguri} for the open string sector. 
As argued in \cite{ooguri}, this expression has a sensible interpretation
in the open string sector for all values of $k$. 

In a  compact CFT, this would be tantamount to a  proof that our 
boundary state is exact.  
However, in the noncompact case Cardy's condition is not as strong, due to 
the fact that there is a continuum of states in the open string spectrum. It
 may be possible then to modify our boundary state at finite $k$ such that 
only 
the density of long string states in affected. This cannot be ruled out based 
on Cardy's condition alone, though we find the scenario implausible.

In \cite{ooguri} a two brane system, located at $\psi_0$ and  $-\psi_0$ was
 also discussed. This system appears not to have such subtleties, and the  
spectrum found in \cite{ooguri} was indeed independent of $\psi_0$. This is in 
complete agreement with our analysis.

\vskip 0.5cm

\section{Acknowledgements}

We are grateful to H.~Liu and J.~Michelson for discussions.
This work was supported in part by  DOE
grant DE-FG02-96ER40559.

\section{Appendix}

$AdS_3$ is the universal cover of the group manifold $\SL$. A convenient
parametrization of a $\SL$ element is provided by the Euler angles:

\bea
g &=& e^{iu\sigma_2} e^{\rho \sigma_3} e^{iv \sigma_2} = \nonumber\\
  &=&
\left(
\begin{array}{lll}
\cos t\cosh\rho+\cos\phi\sinh\rho &\quad & \sin t\cosh\rho-
\sin\phi\sinh\rho\cr
-\sin t\cosh\rho-\sin\phi\sinh\rho &\quad &\cos t\cosh\rho-
\cos\phi\sinh\rho \cr
\end{array}
\right)
\eea
where $\sigma_i, i=1,2,3 $ are the Pauli matrices, and we define:
\beq
u = \frac{1}{2} (t+\phi)~~~~~~~~~~~~ 
v=  \frac{1}{2} (t-\phi)
\eeq

These are the cylindrical (global) coordinates of $AdS_3$. The metric is then:
\beq
ds^2 = - \cosh^2 \rho dt^2 + d\rho^2 + \sinh^2 \rho d\phi^2
\eeq

We will often find it useful to replace the global coordinate system
$(\rho, \phi, t)$ by the coordinates $(\psi, \chi , t)$, with:
\beq
\sinh\psi= \sinh \rho \sin \phi\qquad
\cosh\psi\sinh\chi=-\sinh\rho\cos\phi
\eeq

The $AdS_3$ metric in these coordinates is
\bea
ds^2=d\psi^2+\cosh^2\psi(-\cosh^2\chi dt^2 +d\chi^2)
\eea

Another parametrization of a group element is:

\beq
g  =
\left(
\begin{array}{lll}
X^0+X^1 &\quad & X^2+X^3\cr
X^2-X^3 &\quad & X^0- X^1\cr
\end{array}
\right)
\eeq

with
\beq
\label{hyper}
 (X^0)^2+(X^3)^2-(X^2)^2-(X^1)^2=1
\eeq

This
exhibits $AdS_3$ as a 
3-dimensional hyperboloid embedded in $R^{(2,2)}$. The relation between 
this parametrization
and the global coordinates above  is:

\beq
X^0 +iX^3 = \cosh \rho e^{it} ~~~~~~~~~~~~
X^1 +i X^2 = \sinh \rho e^{i\phi}
\eeq

Euclidean $AdS_3$ is obtained by Wick rotating the global time $t=i\tau$. The metric on the Euclidean $AdS_3$ is therefore:

\beq
ds^2 = \cosh^2 \rho  d\tau^2 + d\rho^2 + \sinh^2 \rho d\phi^2
\eeq

It is sometime  convenient to choose  a different parametrization of
the Euclidean $AdS_3$. In the so-called Poincare coordinates the metric is:

\beq
ds^2 = \frac{1}{r^2}\left( dr^2+ dx^2 +d\tilde{\tau}^2 \right)
\eeq

In the hyperboloid representation of $AdS_3$, the isometry group of the
embedding space $R^{(2,2)}$ is generated by the currents:

\beq
J^{ij}= X^i \frac{\partial}{\partial X^j} - X^j \frac {\partial}{\partial X^i}
\eeq

The  $\SL\times \SL$ generators are then:
\bea
J_1 = \frac{1}{2}(J_{01}+J_{23})~~~~~~
J_2 = \frac{1}{2}(J_{02}-J_{13})~~~~~~
J_0 = \frac{1}{2}(J_{12}+J_{03})\nonumber \\
\bar{J_1} = \frac{1}{2}(J_{01}-J_{23}) ~~~~~~
\bar{J_2} = \frac{1}{2}(J_{02}+J_{13}) ~~~~~~
\bar{J_0} = \frac{1}{2}(J_{12}-J_{03})
\eea

Each set of these generators satisfy the $\SL$ algebra:

\beq
\left[J_1,J_2 \right] = -J_0 ~~~~~~~~~~~
\left[J_1,J_0 \right] = J_2  ~~~~~~~~~~~
\left[J_2,J_0 \right] = J_1            
\eeq

Since these isometries preserve the hyperboloid equation (\ref{hyper}), 
they are
 also isometries of $AdS_3$. For later use it will prove convenient to use 
the generators:
\beq
J^3= iJ_1 ~~~~~~~~~
J^+= -J_2+iJ_3  ~~~~~~~~~~
J^- = J_2 +iJ_3   
\eeq
which satisfy:
\beq
\left[J^0,J^\pm\right] = \pm J^\pm ~~~~~~
\left[J^+,J^- \right] = -2J^3
\eeq
The quadratic Casimir is then:
\beq
J^2 = \frac{1}{2}(J^+J^-  + J^-J^+) -(J^3)^2
\eeq
In global coordinates these generators become 
\bea
J^0&=& \frac{i}{2} \partial_u \nonumber \\
J^+ &=&\frac{i}{2} e^{-2iu}\left[ \coth 2\rho \partial_u-
\frac{1}{\sinh 2\rho } \partial_v + i \partial_\rho \right] \nonumber\\
J^-&=&
\frac{i}{2} e^{2iu}\left[ \coth 2\rho \partial_u-
\frac{1}{\sinh 2\rho }\partial_v- i \partial_\rho \right]
\eea
The generators of the other $\SL$ algebra are obtained by exchanging
$u$ and $v$  in the above expressions.

\end{document}